\documentclass[10pt,aps,prb,superscriptaddress,twocolumn,a4paper,floatfix, nobibnotes]{revtex4-1}
\usepackage{amsmath}%
\usepackage{amsfonts}%
\usepackage{amssymb}%
\usepackage[latin1]{inputenc}
\usepackage{graphicx}
\usepackage{color}
\usepackage[english]{babel}
\usepackage{natbib}
\usepackage[colorlinks=true, citecolor=blue, linkcolor=blue, urlcolor=blue]{hyperref}
\usepackage{booktabs}
\usepackage{enumitem}
\usepackage{subfigure}
\usepackage{multirow}
\usepackage{braket}
\usepackage{dcolumn}
\usepackage{bm}
\usepackage[parse-numbers=true]{siunitx}
\usepackage[geometry]{ifsym}

\begin{document}
\title{Incipient antiferromagnetism in the Eu-doped topological insulator Bi$_2$Te$_3$}
\author{A.~Tcakaev}
\affiliation{Experimentelle Physik IV and R\"{o}ntgen Research Center for Complex Materials (RCCM),
	Fakult\"{a}t f\"{u}r Physik und Astronomie, Universit\"{a}t W\"{u}rzburg, Am Hubland, D-97074 W\"{u}rzburg, Germany}

\author{V.~B.~Zabolotnyy}
\email[Corresponding address: ]{volodymyr.zabolotnyy@physik.uni-wuerzburg.de}
\affiliation{Experimentelle Physik IV and R\"{o}ntgen Research Center for Complex Materials (RCCM),
	Fakult\"{a}t f\"{u}r Physik und Astronomie, Universit\"{a}t W\"{u}rzburg, Am Hubland, D-97074 W\"{u}rzburg, Germany}
	
\author{C.~I.~Fornari}
\affiliation{Experimentelle Physik VII and R\"{o}ntgen Research Center for Complex Materials (RCCM), 
Fakult\"{a}t f\"{u}r Physik und Astronomie, Universit\"{a}t W\"{u}rzburg, Am Hubland, D-97074 W\"{u}rzburg, Germany}
\affiliation{W\"{u}rzburg-Dresden Cluster of Excellence ct.qmat, Germany}	
\affiliation{Laborat\'{o}rio Associado de Sensores e Materiais, Instituto Nacional de Pesquisas Espaciais, S\~{a}o Jos\'{e} dos Campos, 12245-970, S\~{a}o Paulo, Brazil}

\author{P.~R\"{u}{\ss}mann}
\affiliation{Peter Gr\"{u}nberg Institut and Institute for Advanced Simulation, Forschungszentrum J\"{u}lich and JARA, D-52425 J\"{u}lich, Germany}

\author{T.~R.~F.~Peixoto}
\affiliation{Experimentelle Physik VII and R\"{o}ntgen Research Center for Complex Materials (RCCM),
	Fakult\"{a}t f\"{u}r Physik und Astronomie, Universit\"{a}t W\"{u}rzburg, Am Hubland, D-97074 W\"{u}rzburg, Germany}
\affiliation{W\"{u}rzburg-Dresden Cluster of Excellence ct.qmat, Germany}	

\author{F.~Stier}
\affiliation{Experimentelle Physik IV and R\"{o}ntgen Research Center for Complex Materials (RCCM),
	Fakult\"{a}t f\"{u}r Physik und Astronomie, Universit\"{a}t W\"{u}rzburg, Am Hubland, D-97074 W\"{u}rzburg, Germany}	
		
\author{M.~Dettbarn}
\affiliation{Experimentelle Physik IV and R\"{o}ntgen Research Center for Complex Materials (RCCM),
	Fakult\"{a}t f\"{u}r Physik und Astronomie, Universit\"{a}t W\"{u}rzburg, Am Hubland, D-97074 W\"{u}rzburg, Germany}
		
\author{P.~Kagerer}
\affiliation{Experimentelle Physik VII and R\"{o}ntgen Research Center for Complex Materials (RCCM),
	Fakult\"{a}t f\"{u}r Physik und Astronomie, Universit\"{a}t W\"{u}rzburg, Am Hubland, D-97074 W\"{u}rzburg, Germany}	\affiliation{W\"{u}rzburg-Dresden Cluster of Excellence ct.qmat, Germany}	
	
\author{E.~Weschke}
\affiliation{Helmholtz-Zentrum Berlin f\"{u}r Materialien und Energie, Albert-Einstein-Stra\ss{}e 15, D-12489 Berlin, Germany}
\author{E.~Schierle}
\affiliation{Helmholtz-Zentrum Berlin f\"{u}r Materialien und Energie, Albert-Einstein-Stra\ss{}e 15, D-12489 Berlin, Germany}

\author{P.~Bencok}
\affiliation{Diamond Light Source, Didcot OX11 0DE, United Kingdom}

\author{P.~H.~O.~Rappl}
\affiliation{Laborat\'{o}rio Associado de Sensores e Materiais, Instituto Nacional de Pesquisas Espaciais, S\~{a}o Jos\'{e} dos Campos, 12245-970, S\~{a}o Paulo, Brazil}

\author{E.~Abramof}
\affiliation{Laborat\'{o}rio Associado de Sensores e Materiais, Instituto Nacional de Pesquisas Espaciais, S\~{a}o Jos\'{e} dos Campos, 12245-970, S\~{a}o Paulo, Brazil}
	
\author{H.~Bentmann}
\affiliation{Experimentelle Physik VII and R\"{o}ntgen Research Center for Complex Materials (RCCM),
	Fakult\"{a}t f\"{u}r Physik und Astronomie, Universit\"{a}t W\"{u}rzburg, Am Hubland, D-97074 W\"{u}rzburg, Germany}
\affiliation{W\"{u}rzburg-Dresden Cluster of Excellence ct.qmat, Germany}	

\author{E.~Goering}
\affiliation{Max-Planck-Institute for Intelligent Systems, Heisenbergstra{\ss}e 3, 70569 Stuttgart, Germany}

\author{F.~Reinert}
\affiliation{Experimentelle Physik VII and R\"{o}ntgen Research Center for Complex Materials (RCCM),
	Fakult\"{a}t f\"{u}r Physik und Astronomie, Universit\"{a}t W\"{u}rzburg, Am Hubland, D-97074 W\"{u}rzburg, Germany}
\affiliation{W\"{u}rzburg-Dresden Cluster of Excellence ct.qmat, Germany}
\author{V.~Hinkov}
\email[Corresponding address: ]{hinkov@physik.uni-wuerzburg.de}
\affiliation{Experimentelle Physik IV and R\"{o}ntgen Research Center for Complex Materials (RCCM),
	Fakult\"{a}t f\"{u}r Physik und Astronomie, Universit\"{a}t W\"{u}rzburg, Am Hubland, D-97074 W\"{u}rzburg, Germany}

\date{\today}

\renewcommand{\figurename}{FIG.}

\begin{abstract}
Rare earth ions typically exhibit larger magnetic moments than transition metal ions and thus promise the opening of a wider exchange gap in the Dirac surface states of topological insulators. Yet, in a recent photoemission study of  Eu-doped Bi$_2$Te$_3$ films, the spectra remained gapless down to $T=20$ K. Here, we scrutinize whether the conditions for a substantial gap formation in this system are present by combining spectroscopic and bulk characterization methods with theoretical calculations.  For all studied Eu doping concentrations, our atomic multiplet analysis of the $M_{4,5}$ x-ray absorption and magnetic circular dichroism spectra reveals a Eu$^{2+}$ valence and confirms a large magnetic moment, consistent with a $4f^7$ $^8S_{7/2}$ ground state. At temperatures below 10 K, bulk magnetometry indicates the onset of antiferromagnetic (AFM) ordering. This is in good agreement with density functional theory, which predicts AFM interactions between the Eu impurities. Our results support the notion that antiferromagnetism can coexist with topological surface states in rare-earth doped Bi$_2$Te$_3$ and call for spectroscopic studies in the kelvin range to look for novel quantum phenomena such as the quantum anomalous Hall effect.
\end{abstract}

\maketitle

\section{Introduction}

In a magnetic topological insulator (TI), a topologically nontrivial electronic band structure in combination with magnetic order leads to exotic states of quantum matter, such as quantum anomalous Hall (QAH) insulators \cite{QAHCr, QAHV, KOU201534,PhysRevLett.114.187201,PhysRevB.92.201304}, axion insulators \cite{PhysRevLett.120.056801, PhysRevLett.118.246801} and topological superconductors \cite{He294, Sato_2017}. The QAH effect---which is characterized by dissipationless quantized edge conduction in the absence of external magnetic field and Landau levels formation---remains one of the few topological quantum effects unambiguously observed in recent experiments. This new exotic aspect of condensed matter physics, first experimentally discovered in Cr-doped (Bi,Sb)$_2$Te$_3$ TI \cite{QAHCr} and later in V-doped systems \cite{QAHV}, opens a new avenue for the development of low-dissipation electronics, spintronics, and quantum computation \cite{Annurev_He}. However, the key conditions for realizing the QAH effect in TI---low bulk carrier densities and long-range ferromagnetic (FM) order with out-of-plane easy axis --- can be achieved only at millikelvin temperatures ($<100$ mK).

In the past years, great efforts have been made to raise the temperature at which the QAH effect can be observed. Increasing the temperature at least to a few kelvins would already allow the investigation of this effect with more experimental techniques, which could further advance our understanding of it. Unfortunately, the enhanced FM ordering achieved in V-doped (Bi,Sb)$_2$Te$_3$, with the Curie temperature ($T_\text{C}$) twice as high as that of the Cr-doped sample and about one order of magnitude larger coercivity at the same temperature compared to Cr doping, has little influence on the onset temperature of the QAH effect. Only the recent unconventional doping approaches such as magnetic codoping of (Bi,Sb)$_2$Te$_3$ TI with V and Cr were able to increase the temperature of full quantization to 300 mK \cite{doi:10.1002/adma.201703062}, while spatially modulated magnetic doping further increased the temperature of the fully quantized QAH effect in Cr-doped (Bi,Sb)$_2$Te$_3$ up to 0.5 K~\cite{doi:10.1063/1.4935075}. 

Recently, many reports have been published dealing with samples, in which rare-earth ions (RE) instead of transition metals (TM) were used as dopants in order to benefit from their large magnetic moments \cite{Hesjedal_RErev, Harrison2015, Dy_Duffy, Li_Gd_Arpes, Ho_Figueroa}, which might result in a larger Dirac gap in the topological surface states (TSS) \cite{Chen659}. The large magnetic moment of the RE elements, originating from the unpaired $4f$ electrons \cite{RareEarthMagnetism}, would also allow for a decrease in the doping concentration and thus the number of defects, leading to a more stable QAH effect at a higher temperature. The highest effective magnetic moment of $12.6 \mu_\text{B}$ was observed at 2 K for (Dy$_x$Bi$_{1-x}$)$_2$Te$_3$ with $x=0.023$ \cite{Dy_Harrison_2015}. However, the magnetic moment of the Dy ions was found to be strongly concentration dependent, in contrast to Gd and Ho dopants in Bi$_2$Te$_3$ thin films, possessing an effective magnetic moment of $\sim7 \mu_\text{B}$ (close to the maximum free ion value) and of $\sim5.15 \mu_\text{B}$ (half of the theoretical maximum moment), respectively \cite{Hesjedal_RErev}. Despite these large magnetic moments, most investigations found no long-range FM order down to 2 K and thus no gap opening in the TSS \cite{Li_Gd_Arpes, Gd_Harrison2014,Ho_Harrison_2015}. Only in the case of Dy-doping above a critical doping concentration a sizable gap has been reported in angle-resolved photoemission spectroscopy (ARPES), which appears to persist up to room temperature \cite{Harrison2015}. This gap is observed despite the absence of long-range magnetic order, and could originate from short-range FM fluctuations caused by inhomogeneous doping and aggregation of magnetic dopants into superparamagnetic clusters \cite{Dy_Duffy}, as in the case of Cr-doped Bi$_2$Se$_3$ \cite{PhysRevLett.112.056801}. First principle calculations using density functional theory (DFT) suggest that Eu and Sm ions can introduce stable long-range ferromagnetic order in Bi$_2$Se$_3$ \cite{PhysRevB.94.054113}. This, however, was experimentally confirmed only for Sm ions \cite{doi:10.1002/adma.201501254}.

\begin{figure*}
	\centering
	\includegraphics[width=1.0\textwidth]{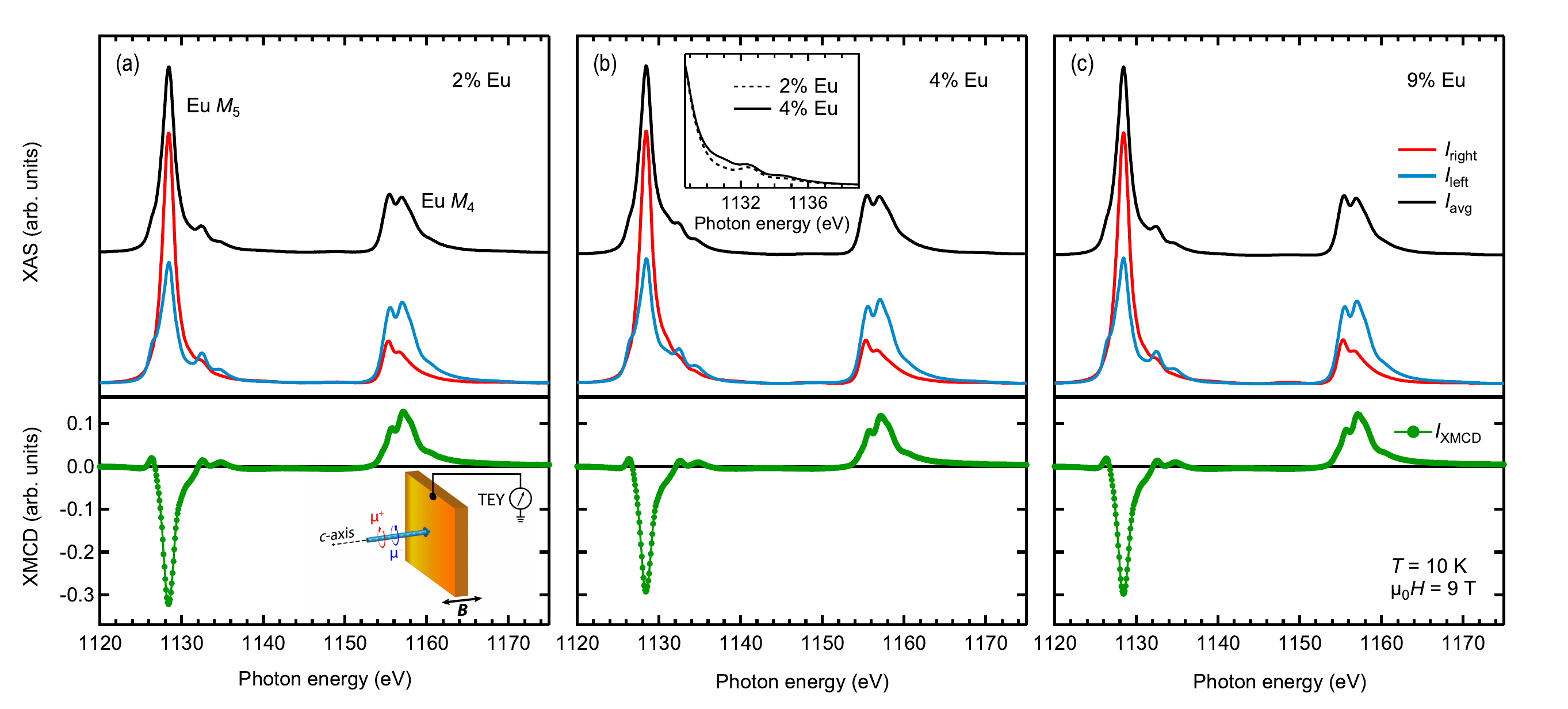}
	\caption{(Color online) Experimental Eu $M_{4,5}$ edges XAS (top) and normalized XMCD (bottom) intensities of (a) 2\%, (b) 4\% and (c) 9\% Eu-doped Bi$_2$Te$_3$ thin films, measured at $T=10$ K in an external magnetic field of 9 T with left circular ($I_{\text{left}}$) and right circular ($I_{\text{right}}$) x-ray polarization. The inset in (a) schematically illustrates the experimental geometry. The inset in (b) highlights the additional small spectral weight at the $M_5$ edge of the 4\% Eu-doped sample, which is absent for the other two samples.}
	\label{fig:XASXMCD}
\end{figure*}

Using antiferro- rather than ferromagnetism has also been studied as an avenue to gapped surface states in layered van der Waals compounds. Recently, the realization of such an antiferromagnetic (AFM) topological insulator in MnBi$_2$Te$_4$ has been reported \cite{Otrokov2019, PhysRevB.100.121104}. It is well known that RE chalcogenides such as EuTe can exhibit AFM order \cite{AFMEUTE, Schierle}. Therefore, it appears promising to take advantage of the larger RE moments to enhance the effect on the TSS in Bi$_2$Te$_3$, just like in the case of FM order.

Whereas MnBi$_2$Te$_4$ is a stoichiometric compound and the AFM order there is intrinsic, here we rather rely on RE doping of Bi$_2$Te$_3$ to induce antiferromagnetism, not least to circumvent RE solubility issues. The general feasibility of this approach has been demonstrated for Ce$_z$Bi$_{2-z}$Te$_3$ \cite{CeBiTe}, Sm$_z$Bi$_{2-z}$Te$_3$ \cite{Jun2020} and Gd$_z$Bi$_{2-z}$Te$_3$ \cite{Kim2015}. As determined by magnetometry \cite{CeBiTe, Jun2020, Kim2015}, the onset of AFM interactions is achieved even at low RE concentrations (in case of Sm $z=0.025$ already suffices). However, x-ray absorption spectroscopy (XAS) and x-ray magnetic circular dichroism (XMCD) investigations addressing the character of the magnetic moments and the impact on the TSS are scarce \cite{Shikin} and limited to temperatures nearly an order of magnitude above the AFM onset temperature, which calls for further investigations.

Here we study Eu$_z$Bi$_{2-z}$Te$_3$ thin films of high structural quality with Eu ions homogeneously incorporated up to a doping level of $z\sim0.2$ \cite{CelsoEudoped}. We provide a comprehensive investigation of a series of samples with three different Eu concentrations. Combining XAS/XMCD obtained at $T\sim10$ K and atomic multiplet calculations allows us to determine the valence state and magnetic moment of the dopants. Using superconducting quantum interference device (SQUID) magnetometry, we observe the onset of antiferromagnetism below about 10~K, which is somewhat unexpected given the prediction of ferromagnetism in the related chalcogenide Eu$_z$Bi$_{2-z}$Se$_3$. Furthermore, we characterize the electronic properties by ARPES and resonant photoemission spectroscopy \mbox{(resPES)} at 20~K. Since this is still above the AFM onset temperature, the TSS remains intact and gapless for all Eu doping levels. Nevertheless, our photoemission measurements allowed us to establish a DFT model, which explains the observed in SQUID data onset of antiferromagnetism by the direct overlap of the wave functions of the Eu impurities.

\section{METHODS}
\label{sec:EXPERIMENTAL}

\subsection{Epitaxial film growth and characterization}

The samples investigated in this work consist of 100~nm thick Eu$_z$Bi$_{2-z}$Te$_3$ films, grown by molecular beam epitaxy on BaF$_2$(111) substrates. The nominal Eu doping concentration is defined as $x_\text{Eu}$ = BEP$_\text{Eu}/$BEP$_{\text{Bi}_2\text{Te}_3}$, where BEP is the beam equivalent pressure of the effusion cells. Four different samples were grown with $x_\text{Eu}$ = 0\%, 2\%, 4\% and 9\%, which would correspond to $z=0, 0.1, 0.2$ and $0.45$ in the chemical formula Eu$_z$Bi$_{2-z}$Te$_3$. Immediately after the growth, all samples were capped by a 100 nm amorphous Te layer to protect the pristine surface from contamination for the x-ray absorption and photoemission measurements. The capping layer was later removed \textit{in situ} right before the spectroscopic measurements \cite{Fornari2016}. The detailed growth conditions and a systematic characterization of the films quality can be found elsewhere \citep{Celso,CelsoEudoped}. X-ray diffraction (XRD) calculations and measurements together with scanning transmission electron microscopy (STEM) images indicate that Eu enters substitutionally on Bi sites up to 4\% of doping, whereas for the 9\% Eu-doped sample EuTe crystalline clusters of 5 to 10 nm are formed \cite{CelsoEudoped}.

\subsection{X-ray absorption spectroscopy}
\label{sec:XASdet}

X-ray absorption spectroscopy (XAS) and x-ray magnetic circular dichroism (XMCD) measurements were carried out using high-field diffractometers  at UE46 PGM-1 beamline, BESSY II, and at beamline I10, Diamond Light Source. Both diffractometers operate under UHV conditions, with a base pressure of $10^{-11}$ mbar. The samples were glued with conducting silver epoxy adhesive onto Cu sample holder and mounted on the cold finger of a helium cryostat. The Te capping layer was mechanically removed ${in\,situ}$ in the fast-entry chamber at a pressure of $10^{-9}$ mbar prior to the measurements. The effectiveness of this method to expose a clean sample surface was demonstrated on Bi$_2$Te$_3$ before \citep{Fornari2016, Tcakaev}.

XAS measurements at Eu $M_{4,5}$ edges were performed at $\sim$10 K and in an external magnetic field of 9 T using circularly polarized light. The degree of circular polarization exceeds $95\%$. The absorption spectra were measured in the total-electron yield (TEY) mode via the sample drain current normalized to the incoming photon intensity ($I_0$). The TEY is known to be surface sensitive, giving a probing depth of 3--6 nm \cite{deGrootBook, DEGROOT1994529}. The XMCD signal was obtained as the difference between two XAS spectra measured in a fixed magnetic field with incoming photons of opposite helicities in normal incidence geometry. The XAS spectra measured with the helicity vector antiparallel (left) and parallel (right) to the fixed magnetic field were scaled with respect to each other to have the same intensity at energies far from the resonances. Using these scaled intensities $I_\text{left}$ and $I_\text{right}$, the average XAS is defined as $I_\text{avg} = (I_\text{left}+I_\text{right})/2$, while the \textit{normalized} XMCD signal is defined as $I_\text{XMCD} = (I_\text{left} - I_\text{right})/(I_\text{left} + I_\text{right})$. Since only the resonant part of the spectra enters the sum rules, the linear background and the continuum edge jumps were subtracted from the raw spectra.

\begin{figure}[t]
	\centering
	\includegraphics[width=1.0\columnwidth]{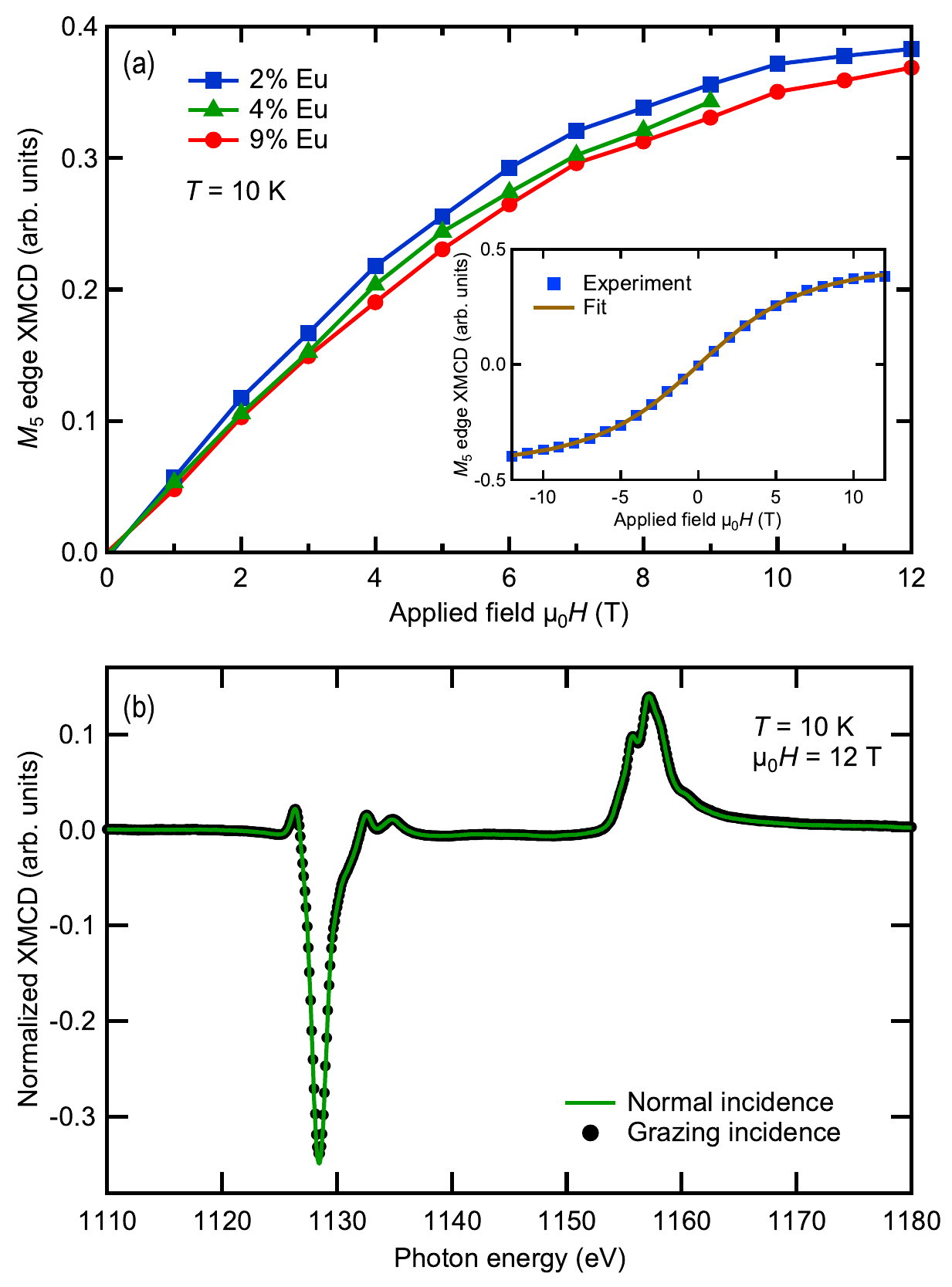}
	\caption{(Color online) (a) Magnetic-field dependence of the $M_{5}$ edge XMCD TEY signal for 2\% Eu-doped (squares), 4\% Eu-doped (triangles) and 9\% Eu-doped (circles) Bi$_2$Te$_3$ measured at $T=10$ K, at normal incidence of the x-rays. The inset exemplarily illustrates the 2\% Eu-doped sample fitted with a Brillouin function (brown line). (b) Normalized XMCD measured at 10 K in an external magnetic field of 12 T for normal incidence of the x-rays, as well as for 70$^\circ$ off-normal, i.e., nearly grazing incidence, showing no noticeable anisotropy.}
	\label{fig:Hyst}
\end{figure}

\subsection{Angle-resolved photoemission spectroscopy}

The photoemission spectra were measured both at laboratory- and synchrotron-based facilities. The laboratory-based angle-resolved photoemission spectroscopy (ARPES) measurements were performed in a UHV system equipped with a Scienta R4000 hemispherical analyzer using He I$_{\alpha}$ radiation  ($h\nu=21.2$ eV). The energy resolution was better than 18 meV and the angular resolution was 0.2$^\circ$. The sample was cooled down to 20 K using a liquid He cryostat. The pressure during the measurement never exceeded $7 \times 10^{-10}$ mbar.

The resonant ($h\nu \sim 1128$ eV) and off-resonant ($h\nu=265$ eV) measurements in the soft x-ray regime were carried out at $T=30$ K using the ASPHERE III end station of the P04 beamline at the PETRA III synchrotron facility (DESY, Hamburg, Germany), with a base pressure better than $2 \times 10^{-10}$ mbar \cite{VIEFHAUS2013151}. 

All studied samples were protected with a Te capping layer, which was removed \textit{in situ} prior to the actual measurement. 

\subsection{Density functional theory calculations}
\label{sec:DFTdet}
Density functional theory (DFT) calculations were performed for Bi$_2$Te$_3$ bulk crystals using the experimental bulk lattice structure \citep{NAKAJIMA1963479} into which Eu defects were embedded.
The electronic structure was calculated within the local spin density approximation \citep{Vosko1980} by employing the full-potential relativistic Korringa-Kohn-Rostoker Green's function method (KKR) \citep{Ebert2011,jukkr} with exact description of the atomic cells \citep{Stefanou1990,Stefanou1991}. The truncation error arising from an $\ell_{max} = 3$ cutoff in the angular momentum expansion was corrected for using Lloyd's formula \citep{Zeller2004}. The Eu defects were embedded self-consistently into the Bi$_2$Te$_3$ crystal using the Dyson equation in the KKR method \citep{Bauer2013} and have been chosen to occupy the substitutional Bi position (denoted by $\mathrm{Eu}_{\mathrm{Bi}}$) in the quintuple layers. We included a charge-screening cluster comprising the first three shells of neighboring atoms and structural relaxations around the defect were neglected. All calculations include spin-orbit coupling self-consistently and were performed for an out-of-plane direction of the magnetic moments of the Eu atoms.
Correlations within the localized $4f$ states of Eu were accounted for using an on-site Coulombic correction (LDA + U)  \citep{Ebert2003} for varying values of the parametrization of $U\in\{0,7,8,9\}\,\mathrm{eV}$ and $J\in\{0,0.75,1.5\}\,\mathrm{eV}$.
To calculate exchange interactions, pairs of Eu impurities were embedded into Bi$_2$Te$_3$ at different distances for substitutional Bi positions within the same quintuple layer. After the self-consistent impurity embedding calculation, the method of infinitesimal rotations \citep{Liechtenstein1987} was used to compute exchange interaction parameters $J_{ij}$ which correspond to the Heisenberg Hamiltonian $\mathcal{H}=\frac{1}{2}\sum_{i,j}\hat{e}_i J_{ij} \hat{e}_j$. Here $\hat{e}_{i}$ indicates the direction of the Eu magnetic moment and $i\neq j$ label the different magnetic Eu atoms. The $J_{ij}$ parameters were calculated using a numerical smearing temperature of 100\,K, which includes the effective contribution of electron scattering due to phonons or intrinsic defects in the Bi$_2$Te$_3$ host crystal that limit the coherence length of the electron's wave functions. Calculations at higher values of the smearing temperature showed a minor effect on the $J_{ij}$'s and are therefore not shown explicitly. 

\subsection{Bulk magnetometry}
The overall magnetic properties of the Eu$_z$Bi$_{2-z}$Te$_3$ films were measured using bulk-sensitive superconducting quantum interference device (SQUID) magnetometry. SQUID measurements were performed as a function of temperature and magnetic field using a 7 T Quantum Design MPMS 3 SQUID VSM. The diamagnetic contribution from the BaF$_2$ substrate was subtracted by high-field linear fitting of M(H) curves  at elevated temperatures (not shown). The temperature dependence of the magnetization was measured in the field-cooled (FC) and zero-field-cooled (ZFC) regimes. In the ZFC measurement, the samples were cooled from room temperature to 2 K without any applied field. After cooling, a magnetic field of 0.1 T was applied perpendicular to the film $c$-axis, i.e., in-plane, and the magnetization was measured upon warming the samples. In the FC measurements the samples were cooled to 2 K in 0.1 T in-plane field and the data were acquired while heating, similar to ZFC.

\begin{figure*}
	\centering
	\includegraphics[width=1.0\textwidth]{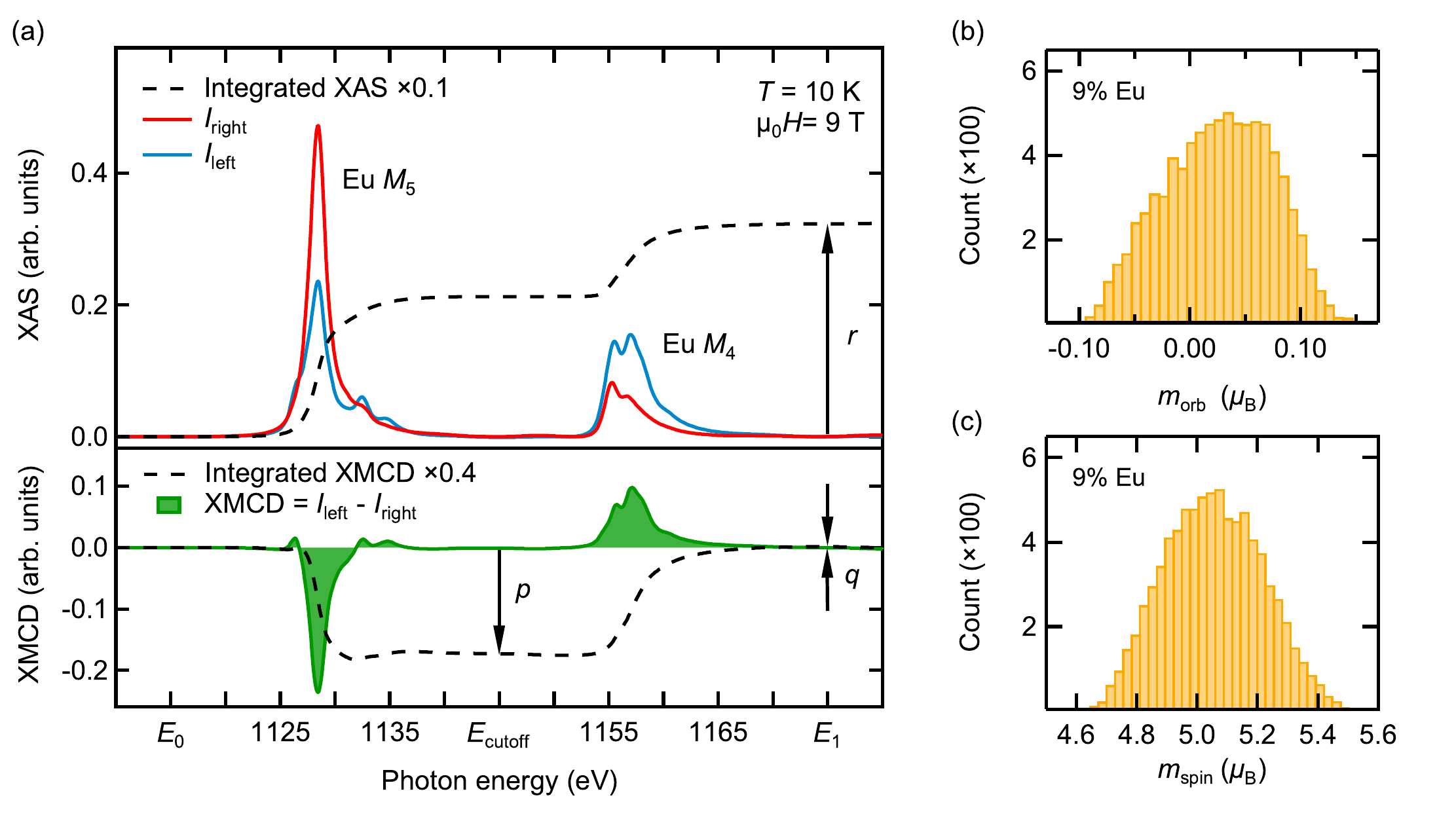}
	\caption{(Color online) Sum rule analysis for the 9\% Eu-doped sample. (a) Left- and right-circularly polarized XAS spectra of the Eu $M_{4,5}$ edges, obtained after the background corrections described in section \ref{sec:EXPERIMENTAL}  ($I_\text{left}$, solid, light blue line, and $I_\text{right}$, solid dark red line), along with the corresponding XMCD data (solid green line, lower panel). The dashed lines show the total integrated XAS and XMCD spectral weight, respectively. The arrows mark the values of $r$, $p$ and $q$ used in Eqs.~(\ref{eqn:OrbbSumRule},~\ref{eqn:SumRule}). $E_0$ and $E_1$ denote the onset and the end energy of the entire $M_{4,5}$ edges, and $E_\text{cutoff}$ denotes the energy separating the $M_4$ and $M_5$ contributions. (b) and (c) Distribution of $m_\text{orb}$ and $m_\text{spin}$, respectively, obtained after application of the sum rules 8192 times, for different sets of fitting parameters, as described in the main text.}
	\label{fig:SumRules}
\end{figure*}

\section{RESULTS and Discussions}
\subsection{Eu $M_{4,5}$ XAS and XMCD}
Fig.~\ref{fig:XASXMCD}  shows XAS and XMCD spectra at the Eu $M_{4,5}$ edges for the 2\%, 4\% and 9\% Eu-doped Bi$_2$Te$_3$ samples. The measurements were conducted at a temperature of $T=10$ K in an applied field of $B=9$ T. The XAS line shapes of all three samples shown in the upper panels are nearly identical and indicate an overwhelming preponderance of Eu$^{2+}$ \cite{PhysRevB.32.5107,doi:10.1143/JPSJ.71.148}. The line shapes of the XMCD spectra shown in the lower panels confirm the Eu$^{2+}$ character, corresponding to a $4f$ electron occupation of $n_f=7$ ($S=7/2$, $L=0$ and $J=7/2$). The small additional spectral weight observed in the 4\% Eu-doped sample (see the inset of Fig.~\ref{fig:XASXMCD}~(b)) probably stems from Eu$^{3+}$, most likely resulting from surface contamination with Eu$_2$O$_3$, as we show in Section~\ref{sec:MLFT} using atomic multiplet calculations. Eu$^{3+}$ is nonmagnetic in the Hund's rule ground state ($S=3$, $L=3$ and $J=0$) and therefore has no contribution to the XMCD spectrum \cite{Wada_1997,Kachkanov}. The electrons of the Eu $4f$ shell are not directly involved in the formation of chemical bonds, unlike the electrons of the $5d$ and $6s$ shells. For this reason, the Eu $M_{4,5}$ absorption spectrum is typically the same for metals, alloys and oxides, apart from small differences in the line shape due to the experimental and lifetime broadening \cite{PhysRevB.32.5107}.  

\begin{figure*}
	\centering
	\includegraphics[width=0.9\textwidth]{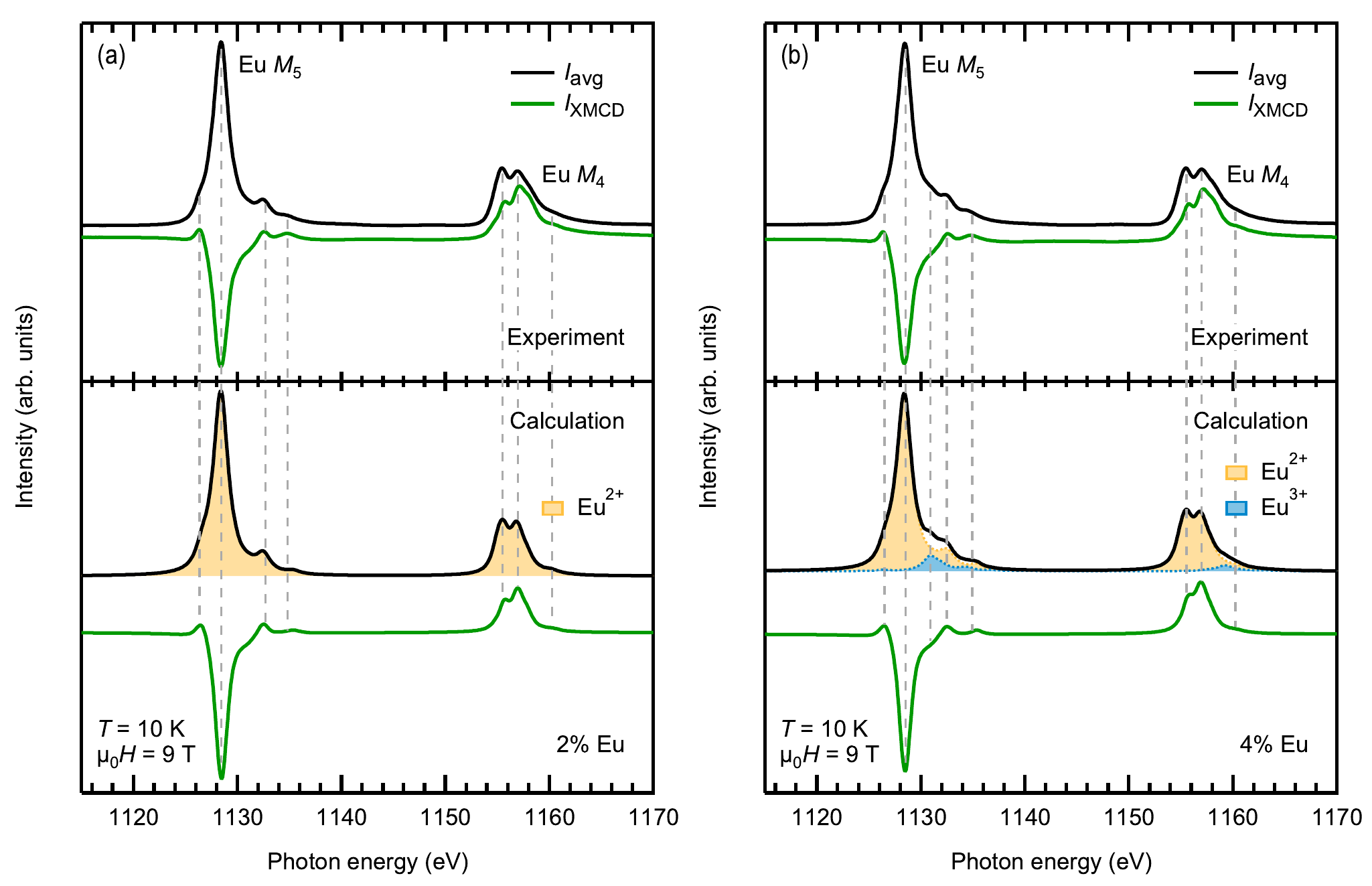}
	\caption{(Color online) Top panel: Experimental Eu $M_{4,5}$ XAS averaged over the two polarizations and XMCD spectra of (a) 2\% and (b) 4\% Eu-doped Bi$_2$Te$_3$ measured at $T=10$ K in an external magnetic field of 9 T. Bottom panel: Calculated average XAS and XMCD spectra for Eu$^{2+}$ and Eu$^{3+}$ obtained by atomic multiplet theory. The dashed vertical lines are drawn as a guide to the eye, highlighting the position of particular features in the spectra.}
	\label{fig:calc4}
\end{figure*}

It is worth mentioning that the strength of the normalized dichroism signal shown in the lower panel of Fig.~\ref{fig:XASXMCD}, which is  directly proportional to the $4f$ magnetic moment of the Eu ion, slightly decreases upon increasing Eu concentration. The same trend was reported for the concentration dependence of the Dy magnetization in Dy:Bi$_2$Te$_3$ films \cite{Dy_Harrison_2015}. The XMCD spectra measured at low temperature in remanence (not shown) display no perceptible response for the entire range of studied concentrations of Eu, thus we observe no evidence for a long-range FM order, also consistent with the SQUID results. 

Previous XAS and XMCD studies of Bi$_2$Te$_3$ thin films doped with RE ions other than Eu revealed a $3+$ valence of the dopants~\cite{Gd_Harrison2014,Dy_Harrison_2015,Dy_FIGUEROA,Ho_Harrison_2015,Dy_Duffy,Ho_Figueroa,Hesjedal_RErev}, in strong contrast with the $2+$ valence of the Eu ions found here. This is likely due to the half-filled $4f$ shell of Eu$^{2+}$, [Xe] $4f^7$, having a very stable Hund's rule ground state ($^8S_{7/2}$) with no spin-orbit splitting and a large spin magnetic moment arising from 7 unpaired electrons. It is also in line with the observation that the trivalent state is the most stable in oxides, while the divalent state is more stable for the  less electronegative chalcogens \cite{CelsoEudoped}. Overall, our XAS and XMCD spectra are in good agreement with those previously reported for Eu $M_{4,5}$ edges \cite{PhysRevMaterials.1.054005,doi:10.1063/1.3076044,PhysRevB.84.235120}. 

\begin{table}[b]
\renewcommand*{\arraystretch}{1.5}
\setlength\tabcolsep{1.6pt}
	\caption{\label{tab:SumRuleAnalysis} Spin, orbital and saturation magnetic moments estimated using XMCD sum rules for Eu $M_{4,5}$ absorption edges (in units of $\mu_\text{B}$). The scaling factor $\mathcal{C}$ was obtained from the fit of the XMCD magnetic field dependence with a Brillouin function, as shown in FIG.~\ref{fig:Hyst}.}
	
	\begin{ruledtabular}
		\begin{tabular}{ccccc}
Sample & $m_\text{spin}$ & $m_\text{orb}$ & $m^\text{sat}_\text{spin}$ & $\mathcal{C}$\\ \hline
$2\%$ Eu & $5.23 \pm 0.21$ & $-0.09 \pm 0.06$ & $6.64 \pm 0.29$ & $1.27 \pm 0.02$\\
$4\%$ Eu & $4.85 \pm 0.20$ & $-0.06 \pm 0.11$ & $6.35 \pm 0.47$ & $1.31 \pm 0.08$\\
$9\%$ Eu & $5.03 \pm 0.22$ & $0.02 \pm 0.07$ & $6.79 \pm 0.39$ & $1.35 \pm 0.05$
		\end{tabular}
	\end{ruledtabular}
\end{table}  

Fig.~\ref{fig:Hyst}(a) illustrates the strength of the XMCD signal as a function of external magnetic field measured at 10 K at normal incidence of the x-rays, revealing the field-dependent magnetization of Eu ions. The data was obtained by sweeping the out-of-plane applied magnetic field in a range of $\pm$12 T at the photon energy of the Eu $M_5$ edge XMCD peak maximum normalized to the off-resonant region. The shapes of the curves are fairly similar for all three Eu concentrations with the XMCD strength slightly decreasing by $\sim6\%$ when going from 2\% to 9\% doping level. No evidence for opening of the hysteresis loop was observed for any of the three samples, which points towards the absence of long-range magnetic ordering of Eu moments. Indeed, the magnetization curves can be closely approximated by a Brillouin function (see Fig.~\ref{fig:Hyst}(a)), which is indicative of paramagnetic behavior. Besides, all magnetization curves are passing directly through the origin, which once again indicates zero remanent magnetization and coercive field. Similar paramagnetic responses were also observed for Gd, Dy and Ho ions doped in thin films of Bi$_2$Te$_3$~\cite{Gd_Harrison2014, Dy_Harrison_2015, Dy_FIGUEROA, Ho_Harrison_2015}. The comparison of XMCD spectra measured at 10 K and external magnetic field of 12 T with normal and grazing x-ray beam incidence is shown in Fig.~\ref{fig:Hyst}(b). No difference between the two spectra can be detected, suggesting no noticeable magnetic anisotropy. 

\begin{table*}
\renewcommand*{\arraystretch}{1.5}
\setlength\tabcolsep{1.6pt}
	\caption{\label{tab:param}Optimized CFT parameters for Eu$^{2+}$ and Eu$^{3+}$ ions used in the atomic multiplet calculation (in units of eV). The best fit yields a reduction of the $F$ and $G$ Slater integrals to 84\% and 74\% of   their Hartree-Fock values, respectively.}
	\label{table:param}
	\begin{ruledtabular}
		\begin{tabular}{cccccccccccccc}
			
			Ion&state&configuration&$F^{(2)}_{ff}$&$F^{(4)}_{ff}$&$F^{(6)}_{ff}$&$\zeta_{4f}$&$F^{(2)}_{df}$&$F^{(4)}_{df}$&G$^{(1)}_{df}$&$G^{(3)}_{df}$&$G^{(5)}_{df}$&$\zeta_{3d}$\\ \hline
			Eu$^{2+}$&initial&$3d^{10}4f^7$&10.913&6.807&4.886&0.160&6.728
			&3.056&4.066&2.379&1.642&11.052
			
			\\
			&final&$3d^{9}4f^8$&11.579&7.238&5.200&0.187&7.347&3.389&4.548&2.664&1.840&11.295

			\\
			Eu$^{3+}$&initial&$3d^{10}4f^6$&$11.826$&$7.422$&5.340&0.175&7.270&3.330&4.446&2.603
			&1.797&11.048\\
			&final&$3d^{9}4f^7$&12.428&7.812&5.624&0.202&7.866&3.656&4.922&2.885&1.993&11.291
			
			\\
		\end{tabular}
	\end{ruledtabular}
\end{table*}

\subsection{Sum rules analysis}

The spin and orbital magnetic moments, which determine the magnetic properties of our thin films, result from the interplay of the hybridization, spin--orbit coupling (SOC), crystal field (CF), Coulomb and exchange interactions. The highly localized and well screened $4f$ electrons of rare earth elements experience comparatively weak crystal fields ($\sim100$ meV) and small hybridizations, with the Coulomb and SOC  interactions being the two dominating energies. Owing to this, RE ions can be considered as exhibiting isolated magnetic moments and, therefore, the materials often show a paramagnetic behavior. 

The magnetic moment of the Eu ion can be readily accessed by means of sum rule analysis. Established by Thole and Carra, the sum rules relate the ratio of integrated XAS and XMCD spectra to the expectation values of spin and orbital angular momenta \cite{Thole_SumRules,Carra_SumRules}. For $3d \rightarrow 4f$ transitions the sum rules are given by 

\begin{equation}
\label{eqn:OrbbSumRule}
\langle L_z \rangle= \bigg(\frac{2q}{r}\bigg) n_h,
\end{equation}
\begin{equation}
\label{eqn:SumRule}
\langle S_z \rangle= - \bigg(\frac{5p - 3q}{2r}\bigg)n_h -3\langle T_z \rangle.
\end{equation}
As indicated in Fig.~\ref{fig:SumRules}(a), $p$ is the integrated intensity of $(I_\text{left} - I_\text{right})$ over the $M_5$ edge, $q$ is the same integral taken over the entire range encompassing the $M_5$ and $M_4$ edges, and $r$ is the intensity of $(I_\text{left} + I_\text{right})$ integrated over the same range as $q$. Furthermore, $n_h$ stands for the number of $4f$ holes and $\langle T_z \rangle$ is the expectation value of the intra-atomic magnetic dipole operator~\cite{Carra_SumRules}. Using the above equations, one can estimate the orbital and spin magnetic moments as $m_\text{orb} = -\langle L_z \rangle \mu_\text{B}$ and $m_\text{spin} = -2\langle S_z \rangle \mu_\text{B}$, respectively. To estimate the required value of $\langle T_z \rangle$ we performed atomic multiplet calculation for Eu$^{2+}$ and found it to be negligibly small $\langle T_z \rangle = -0.004 \hbar$, in good agreement with previously reported values \cite{Tz_vanderLaan}. The number of holes $n_h$ was taken to be 7 for the Eu $4f^7$ valence shell. Similarly to our previous publication, we apply a correction factor to the spin sum rule in order to compensate for the  $jj$ mixing between the $3d_{5/2}$ and $3d_{3/2}$ core levels~\cite{Tcakaev}. However, for Eu$^{2+}$ the correction factor has a rather small value of 1.06,  indicating a low mixing of these two manifolds. 

Since the extracted magnetic moments depend in a nontrivial way on the input parameters controlling the normalization and background subtraction procedures, as well as on the integration energy range ($E_0$, $E_\text{cutoff}$ and $E_1$ shown in Fig.~\ref{fig:SumRules}(a)) and $n_h$, we vary the input parameters in a random and uncorrelated way within the assumed confidence intervals and examine how the final results get distributed, see Fig.~\ref{fig:SumRules}(b, c). In this way we are able to account for possible conjoined effects of the input parameters and produce fair estimates for the uncertainties in $m_\text{spin}$ and $m_\text{orb}$ \cite{Tcakaev}.

Further, we notice that due to the paramagnetic behavior of the Eu magnetization, the external magnetic field of 9 T was  not sufficient to saturate the magnetic moments at $T=10$ K, the temperature at which the data for the sum rule analysis were collected. Therefore we fit the magnetic-field dependence of the $M_5$ edge XMCD signal with a Brillouin function, $B_J(x)$ with $x=\frac{g_JJ\mu_\text{B}B}{k_\text{B}(T-\theta_p)}$ \cite{kittel1996introduction}, as illustrated in the inset of Fig.~\ref{fig:Hyst}(a), which accounts for the finite temperature, and determine the scaling constant $\mathcal{C} = \text{M}(T=0, B=+\infty)/\text{M}(T, B)$. This scaling constant is later used to obtain the magnetic moment at saturation by its value at finite $T$ and $B$. The fit with Brillouin function indicates that to reach 99\% of the full saturation moment at $T=10$~K, one would have to apply an external magnetic field of about 50 T.
 
The results of the sum rules application for the Eu ions are listed in Table~\ref{tab:SumRuleAnalysis}. As expected for Eu$^{2+}$ with its half-filled $4f$ shell, the orbital magnetic moment $m_\text{orb}$ is almost completely quenched for all three concentrations. The values of the saturation spin magnetic moment $m_\text{spin}^\text{sat}$, within the error bars, are also consistent with the $^8S_{7/2}$ ground state for the 2\% and 9\% Eu-doped samples, while for the 4\% doped sample there is some reduction, which could be explained by a non-dichroic contribution coming from the Eu$^{3+}$ contamination. 

In the following section, we will compare the moments obtained with the sum rule analysis with those obtained by atomic multiplet theory.

\begin{figure}[b]
	\centering
	\includegraphics[width=1.0\columnwidth]{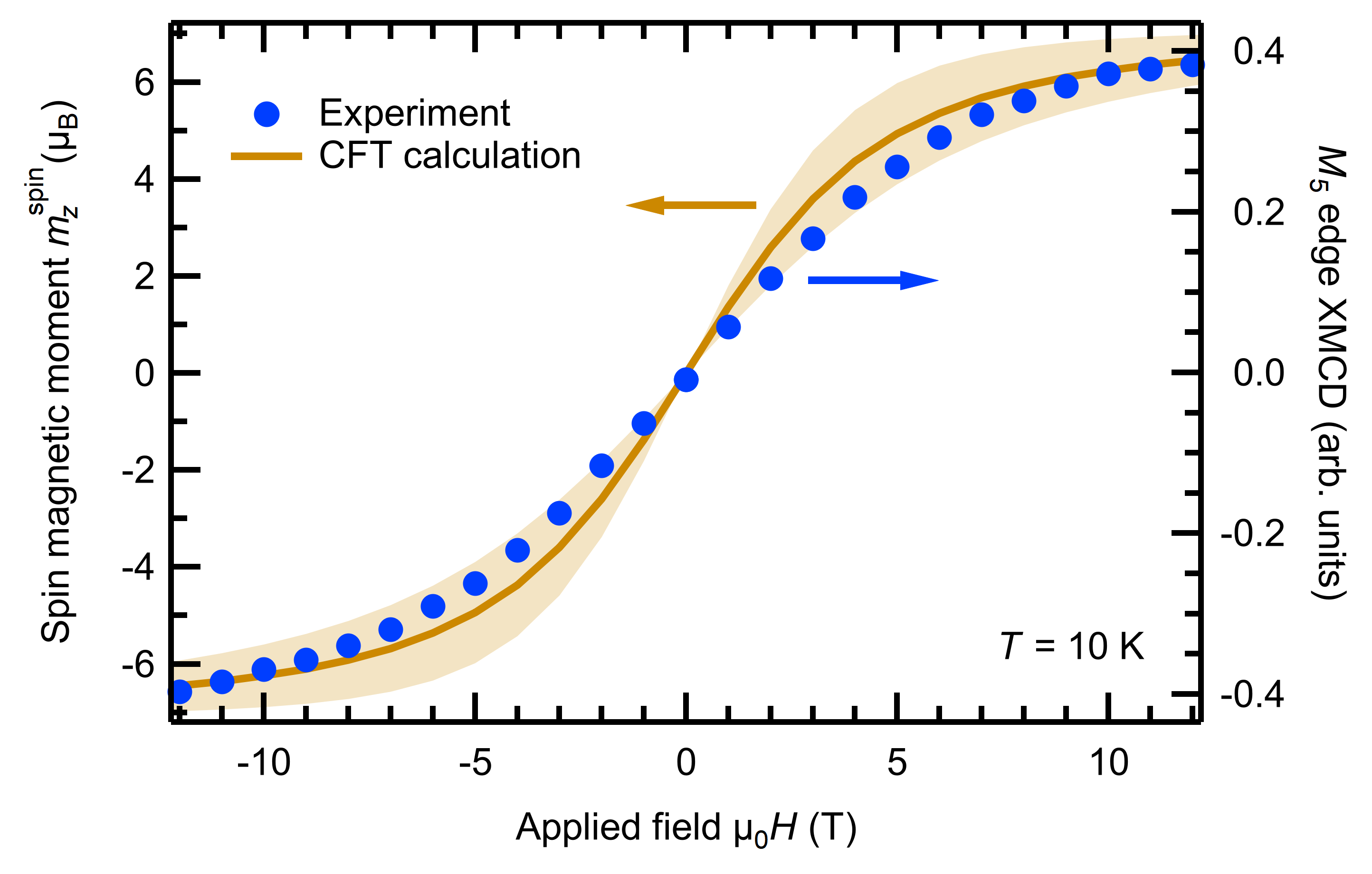}
	\caption{(Color online) Magnetic-field dependence of the CFT calculated spin magnetic moment $m^\text{spin}_z$ (solid line) and of the experimental $M_{5}$ edge XMCD TEY signal for the 2\% Eu-doped Bi$_2$Te$_3$ thin film (full circles) measured at $T=10$ K at normal incidence of the x-rays. The shaded area indicates the error, as estimated in the main text.}
	\label{fig:fielddep}
\end{figure} 

\begin{figure*}
	\centering
	\includegraphics[width=1.0\textwidth]{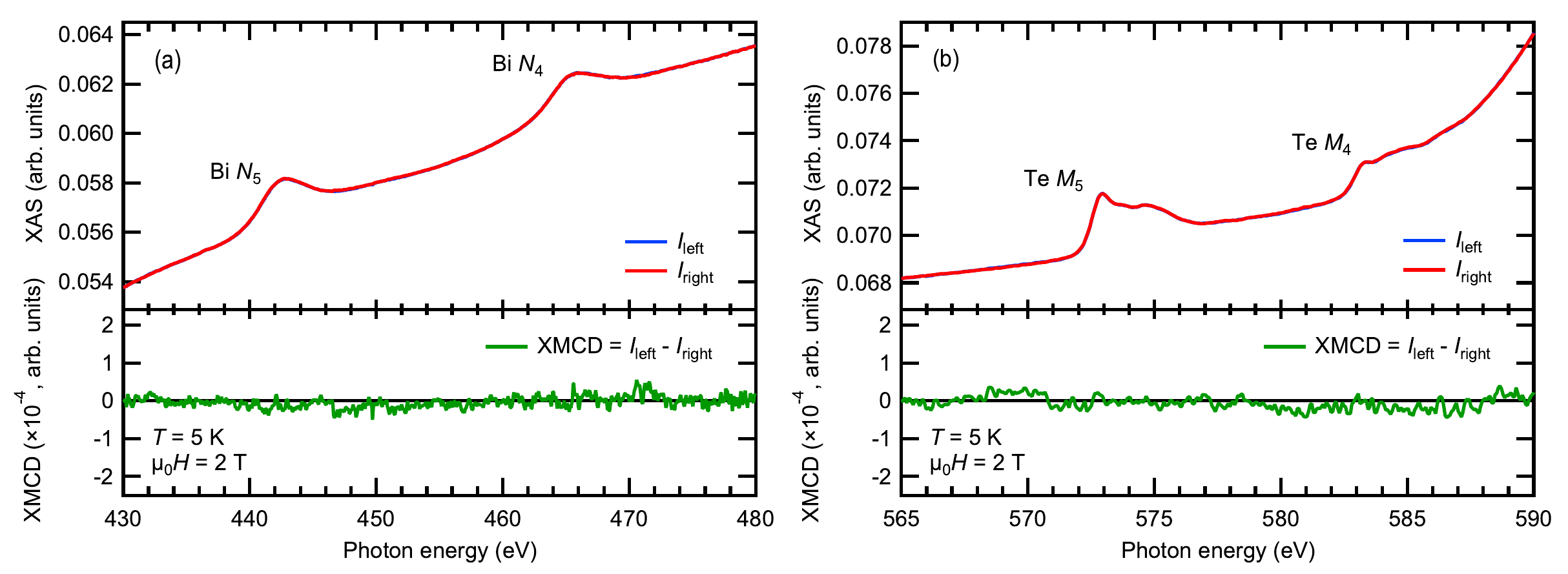}
	\caption{(Color online) Normalized (a) Bi $N_{4,5}$ and (b) Te $M_{4,5}$ XAS (top panel) and XMCD (bottom panel) intensities of the 2\% Eu-doped sample, measured at 5 K in an external magnetic field of 2 T.}
	\label{fig:TeXMCD}
\end{figure*} 

\subsection{Atomic multiplet calculations}
\label{sec:MLFT}

Theoretical XAS and XMCD spectra for the $M_{4,5}$ $(3d \rightarrow 4f)$ absorption of Eu$^{2+}$ and Eu$^{3+}$ ions were calculated using crystal field multiplet theory (CFT) in the framework developed by Thole \textit{et al.} \cite{PhysRevB.32.5107}. The calculation takes into account all the $3d-4f$ and $4f-4f$ electronic Coulomb interactions, as well as the spin-orbit coupling on every open shell of the absorbing atom. The initial values for the Slater integrals were obtained using Cowan's atomic Hartree-Fock (HF) code with relativistic corrections \cite{cowan1981theory}. Their optimized values together with the spin--orbit coupling constants used in the calculations for the Eu$^{2+}$ $3d^{10}4f^7$ and Eu$^{3+}$ $3d^{10}4f^6$ initial state and for the Eu$^{2+}$ $3d^{9}4f^8$ and Eu$^{3+}$ $3d^94f^7$ final state are shown in Table~\ref{tab:param}. The HF values of the direct Slater integrals $F$, determining the size of the electron-electron repulsion, were reduced to 84\%, while those of the exchange Slater integrals $G$ were reduced to 74\%, to account for intra-atomic screening effects~\cite{PhysRevB.32.5107}. These scaling parameters of Slater integrals were found to be the optimal values for the Eu $M_{4,5}$ XAS and XMCD spectra, accurately describing the total spread of the lines in the $3d_{3/2}$ and $3d_{5/2}$ peaks. The strength of the spin--orbit coupling in the $d$-shell was scaled down to 99\% for a better match to the experimental data. The relaxation of atomic orbitals upon the $3d\rightarrow4f$ excitation leads to a slight change in the Slater integrals and the spin--orbit coupling constants $\zeta_{4f}$ and $\zeta_{3d}$. To account for this effect, we used separate sets of these parameters for the initial and final states. As expected, this resulted in a better agreement between the calculated and experimental spectra. The hybridization effect between the localized $f$-electrons and conduction electrons is considered to be weak \cite{PhysRevB.26.2085} and was therefore neglected in the calculations. In our calculation we consider only Eu atoms that substitute Bi in Bi$_2$Te$_3$, which entails $C_{3v}$ symmetry of the CF. Since the nearest 6 Te atoms form almost a perfect octahedron, one could have used $O_h$ symmetry, but we disregard the CF altogether. This simplification is justified by the effective shielding of the external electrostatic potential by the outer $5s$ and $5p$ shells, so the CF splitting in the $f$-shell turns out to be small ($\sim$100 meV) compared to the experimental resolution (120--250 meV) \cite{Shick2015}, and can be neglected in the current consideration. For comparison, in EuO with its divalent state of Eu, the CF value of 175 meV was obtained by means of multiplet calculations of anisotropic x-ray magnetic linear dichroism \cite{PhysRevLett.100.067403}.
 
Calculations were performed using the Quanty software package for quantum many-body calculations, developed by M. W. Haverkort \cite{PhysRevB.85.165113}, which is based on second quantization and the Lanczos recursion method to calculate Green's functions, through avoiding the explicit calculations of the final states. The spectral contributions of the splitted ground state terms to the absorption spectra were weighted using a Boltzmann factor corresponding to the experimental temperature of $T\approx 10$~K. Since the experiments were performed in an external magnetic field of 9 T, this was also included in the calculation. To account for the instrumental and intrinsic lifetime broadening, the calculated spectra were convoluted with a Gaussian function with a standard deviation $\sigma=0.2$~eV and with an energy-dependent Lorentzian profile of 0.4--0.6~eV FWHM. The calculated spectra of Eu$^{2+}$ and Eu$^{3+}$ are linearly superposed with the relative energy position and the relative intensity as adjustable parameters.

Fig.~\ref{fig:calc4} shows the comparison of calculated XAS and XMCD spectra for the 2\% and 4\% Eu-doped samples with experimental data obtained at $T=10$ K and $B=9$ T. We obtain good agreement between experiment and theory, reproducing all essential spectral features and their relative energy positions denoted by vertical dashed lines. This good agreement for the RE $M_{4,5}$ edges is partly due to the CFT being ideally suited to describe transitions into well localized $4f$ states. The calculations for the 2\% and 9\% Eu-doped samples, see Fig.~\ref{fig:calc4}(a), indicate that it is sufficient to consider only divalent Eu to reproduce the experimental spectra with no detectable presence of Eu$^{3+}$. On the other hand, the best fit to the experimental data for the 4\% Eu-doped sample, shown in the lower panel of Fig.~\ref{fig:calc4}(b), is obtained with spectral contributions of 93\% from Eu$^{2+}$ and 7\% from Eu$^{3+}$ ions. In the calculation, the Eu$^{3+}$ spectrum was shifted by 2.5 eV towards higher energies compared to that for the Eu$^{2+}$ state, which is consistent with previous works \cite{Kachkanov,doi:10.1143/JPSJ.71.148,doi:10.1063/1.3251777,PhysRevB.88.024405}. According to the Hund's rules, one would expect a nonmagnetic ground state of Eu$^{3+}$ $^7F_0$ ($S=3$, $L=3$, $J=0$). Due to the nonvanishing interaction with the external magnetic field as compared to the spin--orbit interaction, there is a tiny magnetic moment in the $4f$ shell. However, Eu$^{3+}$ XMCD is much smaller compared to Eu$^{2+}$. The magnetization arising from the Van Vleck paramagnetism of Eu$^{3+}$ due to the admixture of low-lying excited states is also small, with a negligible contribution to the XMCD spectral shape.  

The calculations, which were carried out for Eu$^{2+}$ with the same temperature ($T=10$ K) and external magnetic field ($B=9$ T) as in the experiment, result in a finite orbital moment $m^\text{orb}_z=g_l\langle L_z \rangle=0.02 \mu_\text{B}$, a spin magnetic moment $m^\text{spin}_z=g_s\langle S_z \rangle=6.10 \mu_\text{B}$, and an effective magnetic moment $m^\text{eff}=\sqrt{\langle\mu^2\rangle}=7.91 \mu_\text{B}$. The nonvanishing orbital moment is due to the finite spin--orbit interaction in the $4f$ shell as compared to the Coulomb interaction. As for the Eu$^{3+}$, $m^\text{orb}_z=-0.07 \mu_\text{B}$ and $m^\text{spin}_z=0.15 \mu_\text{B}$. Taking into account the experimental temperature uncertainty, we obtain $m^\text{spin}_z=(6.10 \pm 0.44) \mu_\text{B}$, which is reasonably close to the XMCD sum rules results listed in Table~\ref{tab:SumRuleAnalysis}. Sum rules and atomic multiplet calculations also yield similar results for $m^\text{orb}_z$. Possible causes for the small deviation of the sum rules extracted spin magnetic moments from the multiplet calculations are non-magnetic contributions of the Eu sites or non-collinear alignment of the Eu ions in the paramagnetic phase, as well as partial antiferromagnetic coupling between the Eu ions \cite{PhysRevB.78.134423, AFMEUTE}. 

Fig.~\ref{fig:fielddep} shows the magnetic field dependence of the CFT calculated $m^\text{spin}_z$ for Eu$^{2+}$. Within the error bars resulting from experimental temperature uncertainty, it well reproduces the experimental field-dependent magnetization of Eu ions in Bi$_2$Te$_3$ at $T=10$ K. 

\begin{figure}
	\centering
	\includegraphics[width=1.0\columnwidth]{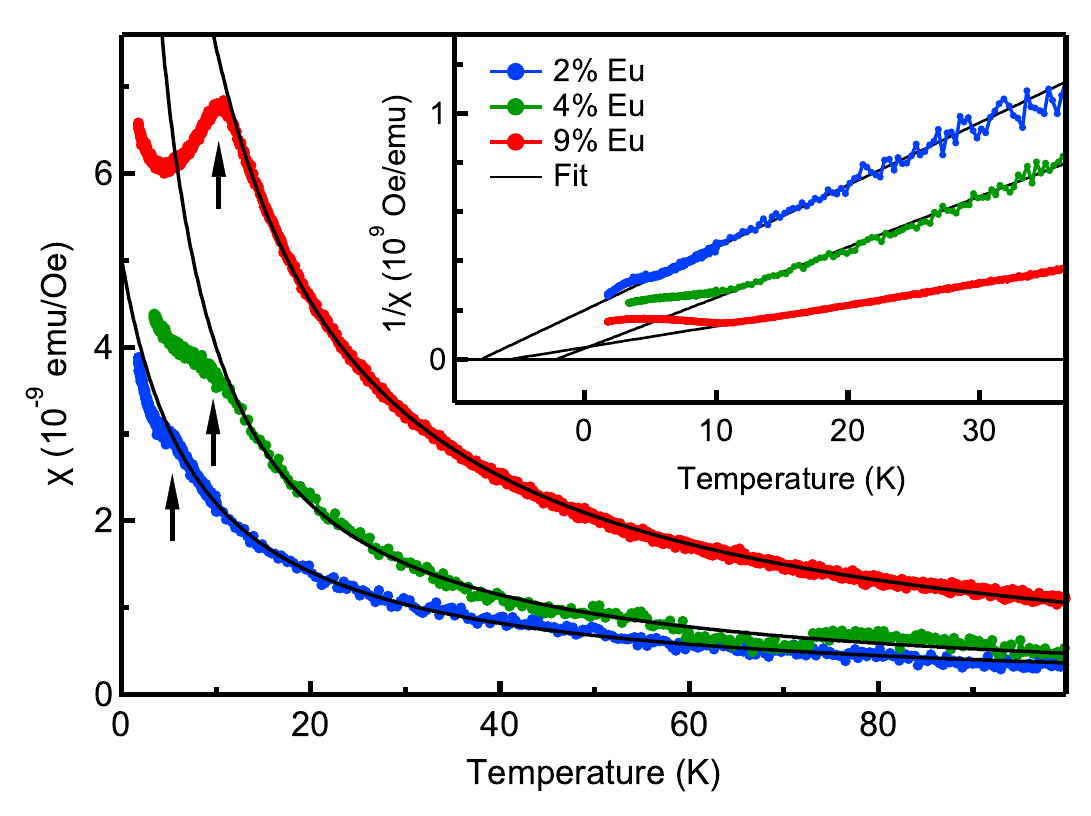}
	\caption{(Color online) Field-cooled magnetic susceptibility $\chi$ for 2\% Eu-doped (solid black), 4\% Eu-doped (solid red) and 9\% Eu-doped (solid blue) Bi$_2$Te$_3$ in a 2--100 K temperature range. The arrows indicate the N\'{e}el temperature estimated from the cusp in the $\chi(T)$ curves. In the inset, a comparison of the inverse magnetic susceptibility for the corresponding samples is shown at low temperatures from 2 K to 35 K. The black solid lines represent linear Curie--Weiss fits to the experimental data.}
	\label{fig:SQUID}
\end{figure}
	
\begin{figure*}
	\centering
	\includegraphics[width=1.0\textwidth]{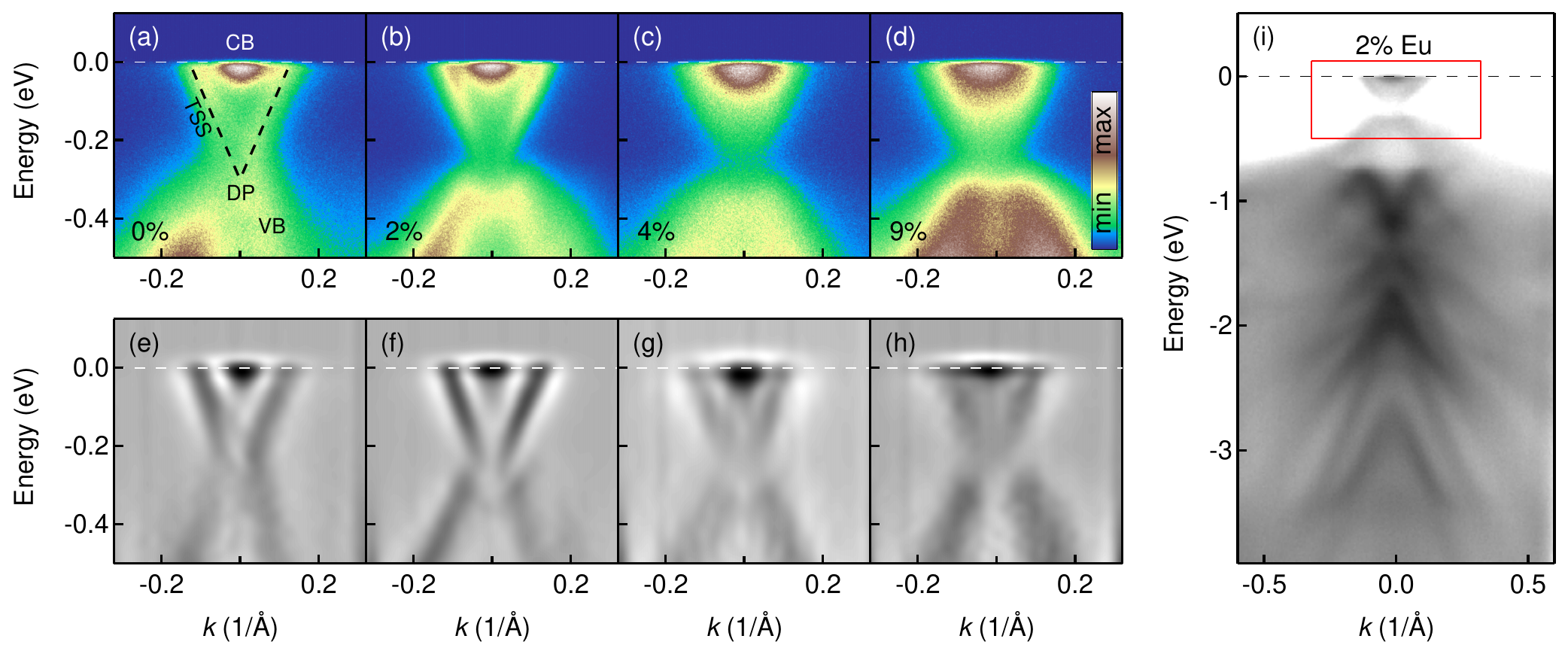}
	\caption{(Color online) (a-d) ARPES spectra of Eu-doped Bi$_2$Te$_3$ thin films with doping ranging from 0\% to 9\%, measured near the $\bar{\Gamma}$ point at 20 K using a photon energy $h\nu = 21.2$~eV. (e-h) Corresponding second derivative plots. (i) Wide energy range spectrum near the $\bar{\Gamma}$ point for the 2\% Eu-doped sample.}
	\label{fig:ARPES}
\end{figure*}
	
\subsection{Te $M_{4,5}$ and Bi $N_{4,5}$ XAS and XMCD}
In our recent work comparing the magnetic properties of V- and Cr-doped (Bi,Sb)$_2$Te$_3$ \cite{Tcakaev}, we demonstrated a significant XMCD signal detected at nominally non-magnetic Sb and Te host atoms due to the strong $pd$-hybridization between TM dopants and the host material. Here, in the case of Eu-doped Bi$_2$Te$_3$, we have also checked for dichroism at Bi and Te sites. Fig.~\ref{fig:TeXMCD} displays the XAS and XMCD measurements at the Bi $N_{4,5}$ and Te $M_{4,5}$ absorption edges at a temperature of 5 K in an applied magnetic field of 2 T. There is no spin polarization detectable on the Te and Bi sites for any dopant concentration. This indicates that a possible magnetic interaction between Eu atoms is not mediated through Te or Bi by means of some sort of indirect exchange.

\subsection{Bulk magnetometry results}
\label{sec:SQUID}

The bulk magnetic properties of our samples were investigated using a laboratory-based SQUID magnetometer. Fig.~\ref{fig:SQUID} shows the field-cooled magnetic susceptibility $\chi$ as a function of temperature for all three samples, measured in an in-plane applied magnetic field of 100 mT. The inset compares the inverse magnetic susceptibilities $1/\chi$ for all three samples as a function of temperature. The inverse magnetic susceptibility data can be fitted using the Curie--Weiss law $\chi=\chi_0+C/(T-\theta_p)$ (shown with solid black lines), where $\chi_0$ represents the temperature-independent contribution, $C$ is the Curie constant, and $\theta_p$ is the Weiss temperature. Fitting the data in the high temperature range reveals a negative Weiss temperatures $\theta_p=-7.8$ K, $-2.1$ K and $-5.6$ K for the 2\%, 4\% and 9\% Eu-doped samples, respectively. These negative values suggest an existence of AFM ordering at low temperatures, below the temperature of about $10$~K at which the XMCD data was acquired. A similar behavior was also reported for Gd-, Dy- and Ho-doped Bi$_2$Te$_3$ thin films~\cite{Hesjedal_RErev}. For the Gd-doped single crystals Gd$_x$Bi$_{2-x}$Te$_3$ with $x = 0.20$, the magnetic phase transition from a PM phase to an AFM phase was reported to occur at the N\'{e}el temperature $T_\text{N}=12$~K \cite{Kim2015}.

The Weiss temperature $\theta_p$ does not always equal the N\'{e}el temperature $T_\text{N}$ \cite{coey_2010, hellwege2013einfuhrung}. However, $T_\text{N}$ can also be estimated by the position of the cusp feature in the temperature dependence of the magnetic susceptibility $\chi(T)$ (see Fig.~\ref{fig:SQUID}). For the 2\% Eu-doped sample we find $T_\text{N}\approx6.0$~K, while for the 4\% and 9\% samples the N\'{e}el temperature is about 9.0~K and 10.5~K. This seems to be the expected simple monotonic behavior as a function of Eu concentration. Increased concentration results in a higher interaction strength due to the shorter average distances between Eu ions, and hence in a higher N\'{e}el temperature. 

As we have previously discussed \cite{CelsoEudoped}, the 9\% sample, stretching the solubility limit of Eu in Bi$_2$Te$_3$, is prone to Eu inhomogeneities and clustering. Therefore it is possible that the much more pronounced cusp feature in the case of the 9\% sample is related to AFM EuTe crystalline clusters. For example, for Eu-doped GeTe bulk crystals, AFM order was observed due to EuTe clusters at $T_\text{N}\approx11$ K \citep{GeEuTe}. In fact, EuTe is a well known magnetic semiconductor and a prototypical Heisenberg antiferromagnet below $T_\text{N}=9.8$~K \citep{Schierle}.

\subsection{Electronic properties}

To study the effect of Eu dopants on the electronic structure of Bi$_2$Te$_3$ we have performed extensive laboratory- and synchrotron-based photoemission measurements.

The laboratory-based angle-resolved spectra (ARPES) were taken at 20 K using He I$_\alpha$ radiation ($h\nu=21.2$ eV) right after mechanical removal of the Te capping layer. In Fig.~\ref{fig:ARPES}~(a-d) we show the data for all samples, including the undoped reference sample. While the M-shaped bulk valence band (VB) and the bulk conduction band (CB) can be seen for all samples, the topological surface state (TSS) is clearly observed only up to 2\% doping. For the higher levels, the spectra are getting blurred because of the increased structural disorder \cite{CelsoEudoped}. To better highlight the bands, we supplement these data with the second derivative plots shown in Fig.~\ref{fig:ARPES}~(e-h) \cite{RevModPhys.75.473, PhysRevB.88.140505}. The gapless TSS can now be seen for all doping levels. The estimated Fermi velocity ranges from 2.55~eV$\cdot$\AA~($3.9\cdot10^5$ m/s) to 2.63~eV$\cdot$\AA~($4.0\cdot10^5$ m/s), which is in excellent agreement with the previous data for undoped bulk samples \citep{Chen2009}.

In Fig.~\ref{fig:ARPES}~(i) we also show a wide energy scan for the 2\% sample. The red rectangle highlights the position of the TSS, the CB and the top of the VB. The VB observed at higher binding energies closely resembles that of the undoped Bi$_2$Te$_3$, with no signatures of Eu impurity bands. Here, though, one should keep in mind that the photoemission matrix elements may cause a drastic intensity variation between different bands. Under unfavorable conditions, this may result in swamping of a weak impurity signal by a more intense feature.

\begin{figure*}
	\centering
	\includegraphics[width=1.0\textwidth]{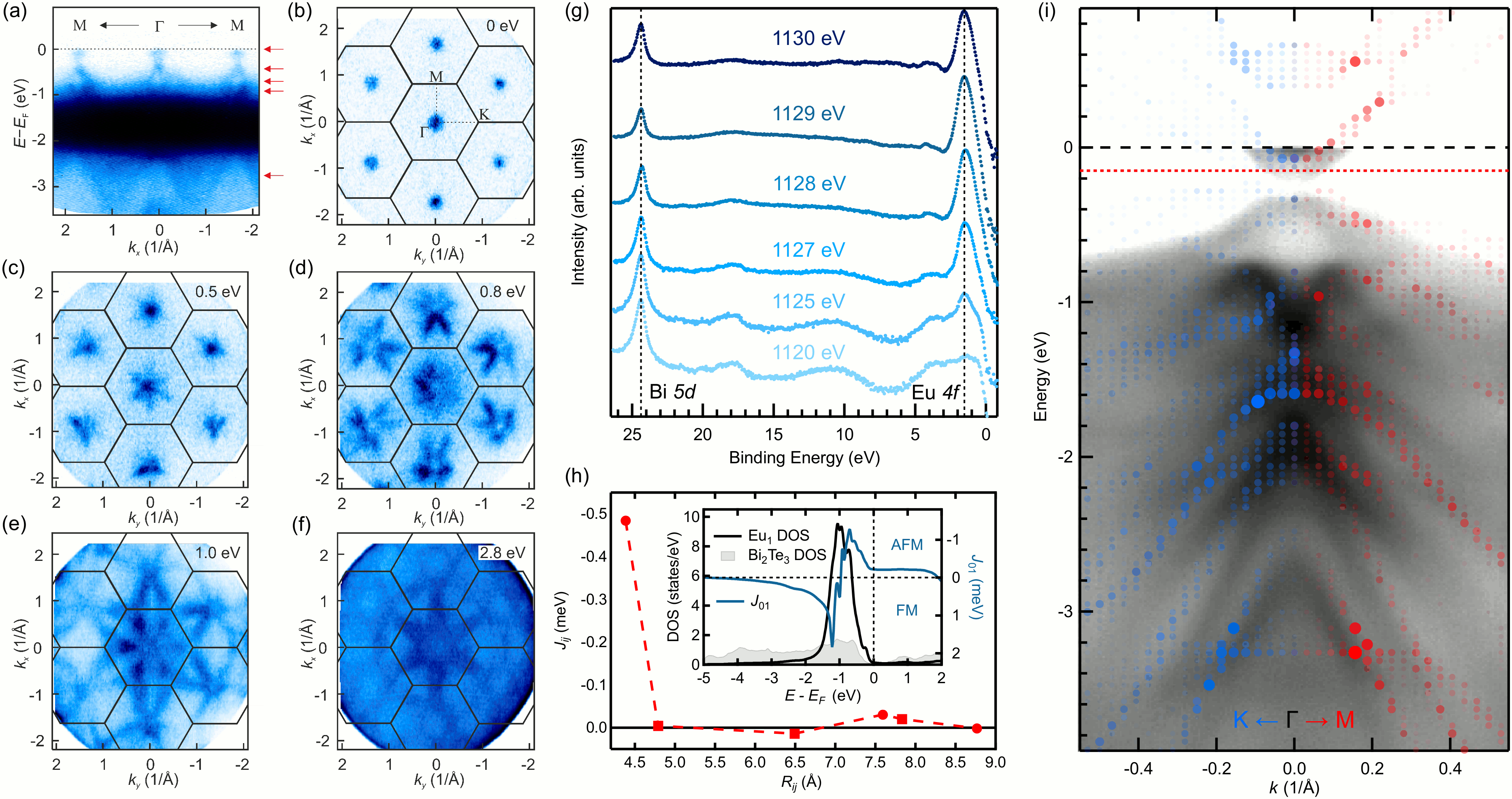}
	\caption{(Color online) (a) Energy-momentum cut along the $\bar{\Gamma}\bar{\text{M}}$ direction for the 2\% sample. (b-f) Constant energy maps for binding energies ranging from 0 to 2.8 eV. The hexagonal shapes are the boundaries of the 2D Brillouin zones. (g) ResPES data for the 4\% sample, showing the Bi $5d$ and Eu $4f$ core levels. (h) Distance dependence of the exchange interaction $J_{ij}$ between pairs of Eu impurities placed in the same ({\large$\bullet$}) or neighboring ({\tiny$\blacksquare$}) Bi layers within one quintuple layer. The inset shows $4f$ dominated total DOS of the Eu dimer (black) and total DOS of the host material (gray) together with the energy dependence of the exchange coupling strength $J_{01}$ (blue). (i) LDA band structure and ARPES data for the 2\% sample. The LDA bands are shown in $\Gamma$M and $\Gamma$K directions with symbols, whose size and opaqueness is proportional to the spectral weight. The red dotted line shows the position of the theoretical Fermi level, which differs from the experimental one by 150 meV due to the intrinsic n-doping in the measured sample.}
	\label{fig:DFT_PES}
\end{figure*}

To conclusively check for the presence of Eu impurity bands we have performed additional synchrotron-based measurements using different excitation energies. In Fig.~\ref{fig:DFT_PES}~(a) we show a $\bar{\Gamma}\bar{\text{M}}$ spectrum taken from the 2\% sample at 10 K using 265 eV photons. Along with the characteristic Dirac cones and the top of the VB, we now clearly see the impurity band located between 1.1 and 2.25 eV. To better illustrate the dispersion of different features, in Fig.~\ref{fig:DFT_PES}~(b-f) we also plot several constant energy maps for binding energies ranging from 0 to 2.8 eV. The hexagonal shapes denote the boundaries of the 2D Brillouin zones. Whereas the features seen at the Fermi level around the $\bar{\Gamma}$ points (Fig.~\ref{fig:DFT_PES}~(a,b)) are due to the TSS and the CB, the band structure at higher binding energies outside the region dominated by the impurity bands (Fig.~\ref{fig:DFT_PES}~(a,c-f)) is very much like that of typical Bi$_2$Te$_3$.

As in the case of V-doped (Bi,Sb)$_2$Te$_3$ \cite{PhysRevB.94.195140}, to confirm that the observed flat feature is indeed the Eu $4f$ impurity band we have performed resPES measurements. Fig.~\ref{fig:DFT_PES} (g) shows the resPES spectra for the 4\% doped sample taken with a photon energy ranging from 1120 eV to 1130 eV. For photon energies below the resonant one ($h\nu < 1120$ eV) only the Bi $5d$ core level and the valence band are visible. By gradually tuning the photon energy to the Eu $3d\rightarrow4f$ resonance, we see a peak growing around 1.7 eV binding energy. The intensity increase is more than hundredfold, which eventually confirms that the observed feature is the Eu $4f$ impurity band.

\subsection{Magnetic exchange coupling calculation}
In order to understand the magnetic properties of Eu-doped Bi$_2$Te$_3$ and the magnetic exchange coupling mechanism between Eu ions, we have performed DFT calculations for bulk-doped Bi$_2$Te$_3$ including the effect of correlations within the LDA+U method, as outlined in section \ref{sec:DFTdet}. The calculated electronic band structure of the host material overlaid with experimental ARPES data is shown in Fig.~\ref{fig:DFT_PES}~(i). The band dispersion in the $\Gamma$M (red dots) and $\Gamma$K (blue dots) directions shows good agreement with the experimental data.

We find that the occupied $4f$ states of $\mathrm{Eu}_{\mathrm{Bi}}$ exhibit a rigid shift down in energy with increasing $U^{\mathrm{eff}}=U-J$, which changes the magnetic moment from $6.58\mu_B$ in the case of pure LDA ($U=0, J=0$) to $6.94\mu_B$ in case of LDA+U ($U^{\mathrm{eff}}=8.25\,\mathrm{eV}$). The size of the magnetic moment and the bandwidth of the $4f$ states do not change much for $U^{\mathrm{eff}}$  values ranging from 4.25 to 9 eV, which indicates a stable half-filling of the Eu $4f$ orbitals and only a weak $pf$-hybridization with the Bi$_2$Te$_3$ host system. This is consistent with the very small and antiferromagnetically aligned induced magnetic moments in the first two Te and the first Bi neighbors around the $\mathrm{Eu}_{\mathrm{Bi}}$ defect of $-8.0\times10^{-3}\,\mu_\text{B}$, $-4.3\times10^{-3}\,\mu_\text{B}$ and $-6.9\times10^{-5}\,\mu_\text{B}$, respectively. Additionally, we performed band structure calculations in a $2\times2\times1$ supercell, which show that the $\mathrm{Eu}_{\mathrm{Bi}}$ impurity bands do not disperse due to the weak hybridization of the $f$ states with its surrounding $p$ states. This result agrees with the results of the ARPES measurements, see Fig.~\ref{fig:DFT_PES}~(a).

In Fig.~\ref{fig:DFT_PES}~(h) we show the distance dependent exchange coupling constants $J_{ij}$ for $U^{\mathrm{eff}}=8.25\,\mathrm{eV}$. The inset shows the impurity DOS and the energy-dependent exchange coupling $J_{01}$ for nearest $\mathrm{Eu}_{\mathrm{Bi}}$-$\mathrm{Eu}_{\mathrm{Bi}}$ neighbors ($R_{ij}=4.38$\,\AA) which is in good agreement with the experimentally determined position of the Eu $4f$ states. We find weak antiferromagnetic interactions for the first-neighbor $\mathrm{Eu}_{\mathrm{Bi}}$ impurities that are located on the same Bi layer. For larger distance between the Eu atoms the exchange interactions quickly decline. The energy-resolved $J_{ij}$ reveals a flat plateau of antiferromagnetic interactions above the Fermi level, which increases for smaller $U^{\mathrm{eff}}$ (not shown). This indicates antiferromagnetic coupling arising from the direct overlap of the impurity wave functions \citep{Belhadji2007, Ruessmann2018}. The strong spatial localization of the Eu $4f$ states explains the weakness of the interaction and the quick decrease with distance. 

Additional calculations with a Fermi level shifted into the bulk conduction band show that the strength of the antiferromagnetic exchange interactions can be increased by up to $\approx50\%$ for Fermi level shifts of up to $\pm0.4\,\mathrm{eV}$. However, the weak $pf$-hybridization between Eu impurity and surrounding host atoms does not result in a significant increase of exchange interactions at larger distances.

\section{Conclusions and Outlook}
\label{sec:SUMMARY}

Realizing an antiferromagnetic topological insulator by doping Bi$_2$Te$_3$ with Eu has turned out to be more challenging than realizing its FM counterpart, namely V or Cr doped Bi$_2$Te$_3$. One likely reason is the adverse effect of the random---and dilute---impurity distribution on establishing a staggered magnetization.

The disorder and charge doping induced by the non-isoelectronic substitution present another challenge as they can interfere with the integrity of the TSS. Our comprehensive experimental and theoretical studies indicate that Eu$_z$Bi$_{2-z}$Te$_3$ is not critically affected by these problems. First, the TSS remain detectable in our ARPES results at all Eu concentrations. This is noteworthy since Eu, unlike most other RE elements, enters Bi$_2$Te$_3$ as Eu$^{2+}$ and thus leads to hole doping and disorder \cite{CelsoEudoped}.

Second, for all Eu concentrations our SQUID data yield a negative Weiss temperature $\theta_p$ and a cusp-like feature in the $\chi(T)$ curve, which indicates the onset of antiferromagnetic order for temperatures between 5~K and 10~K. Due to the thinness of the samples and the presence of Eu, it was not possible to measure the antiferromagnetic correlation length experimentally using neutron diffraction. However, the antiferromagnetic coupling between Eu atoms is corroborated by our DFT calculations in the LDA+U approximation, which well reproduce our photoemission data. The largest effective $J_{ij} = -0.5$ meV is found between Eu ions inside the same Bi layer with its energy comparable to the AFM onset temperature observed in the SQUID data. We point out that previous theoretical studies of Eu$_z$Bi$_{2-z}$Se$_3$ predicted FM order \cite{PhysRevB.94.054113}, for which we found no evidence in our related telluride system.

Considering the hexagonal arrangement of the atoms in the Bi layer, one would expect the AFM order to get stronger with increasing Eu doping, but then at higher levels increasing frustration should suppress ordering. Counter to this intuition, our SQUID data seem to indicate an increasing AFM onset temperature up to the $9\%$ doping, for which nearly one out of four Bi atoms is replaced by Eu. This is probably due to exceeded Eu solubility in the $9\%$ sample and cluster formation of EuTe \cite{CelsoEudoped}, which is a well known antiferromagnet with $T_\text{N}=9.8~\text{K}$.

Whereas in MnBi$_2$Te$_4$ the interactions are ferromagnetic within the Bi planes and antiferromagnetic between the neighboring planes \cite{Otrokov2019, Wueaax9989}, our theoretical prediction of antiferromagnetism in Eu$_z$Bi$_{2-z}$Te$_3$ is different: It closely resembles that found in Gd$_z$Bi$_{2-z}$Te$_3$, for which DFT calculations yield AFM coupling between Gd atoms inside a Bi plane \cite{Kim2015}, just like for the Eu atoms in our case. In addition, a gap formation was experimentally observed for Gd doping \cite{Shikin}, but its connection to the in-plane AFM interactions still needs to be clarified.

In conclusion, our results warrant further investigations at temperatures below 10~K down to the kelvin range to better understand the character of the antiferromagnetism we observe and to experimentally establish its impact on the TSS. Kelvin-range photoemission and XAS experiments are challenging and were not performed for the present study. Yet, in the light of our results, XAS and dichroism measurements, including linear dichroism to characterize the AFM state, appear promising. Low-temperature ARPES needs to be performed to search for a gap opening in the TSS.

The onset of antiferromagnetism over a substantial doping range corroborates the potential of RE doping to result in an AFM topological insulator with exotic quantum properties.

\section{Acknowledgments}
This work was funded by the Deutsche Forschungsgemeinschaft (DFG, German Research Foundation) -- Project-ID 258499086 -- SFB 1170 (projects C04, C06 and A01), the W{\"u}rzburg-Dresden Cluster of Excellence on Complexity and Topology in Quantum Matter -- ct.qmat (EXC 2147, Project-ID 390858490) and the BMBF (Project-ID 05K19WW2). This work was also supported by FAPESP (grant No. 2016/22366-5). We acknowledge Diamond Light Source for time on Beamline I10 under Proposal MM19994. We also thankfully acknowledge HZB for the allocation of synchrotron radiation beamtime and the financial support. Parts of this research was done at PETRA III (DESY, Hamburg, Germany) under Proposal No. I-20181060. We would like to thank F. Diekmann, S. Rohlf, M. Kall{\"a}ne and the staff of beamline P04 for experimental support. C.I.F. acknowledges the Hallwachs-R{\"o}ntgen Postdoc Program of ct.qmat for financial support. PR acknowledge support by the Deutsche Forschungsgemeinschaft (DFG, German Research Foundation) under Germany's Excellence Strategy Cluster of Excellence Matter and Light for Quantum Computing (ML4Q) EXC2004/1 390534769. This work was supported by computing time granted by the JARA Vergabegremium and provided on the JARA Partition part of the supercomputer CLAIX at RWTH Aachen University.


%

\end{document}